\documentclass[aps,prl,reprint,superscriptaddress,footinbib]{revtex4-1}

\usepackage{graphicx}% Include figure files

\begin{document}

%\preprint{}

\title{The Spatial Crossover between Far-From-Equilibrium and Near-Equilibrium Dynamics in Locally Driven Suspensions}


\author{Ilya Svetlizky}
\affiliation{School of Chemistry, Tel-Aviv University, Tel-Aviv 6997801, Israel}
\affiliation{School of Engineering and Applied Sciences, Harvard University, Cambridge, Massachusetts 02138, USA}
\author{Yael Roichman}
\affiliation{School of Chemistry, Tel-Aviv University, Tel-Aviv 6997801, Israel}
\affiliation{School of Physics and Astronomy, Tel-Aviv University, Tel-Aviv 6997801, Israel}

\date{\today}

\begin{abstract}
We examine the response of a quasi-two-dimensional colloidal suspension to a localized circular driving induced by optical tweezers. This approach, a microscale version of rheometry, allows us to resolve over three orders of magnitude in P\'eclet number ($Pe$) and provide a direct observation of a sharp spatial crossover from far- to near-thermal-equilibrium regions of the suspension. In particular, particles migrate from high to low $Pe$ regions and form strongly inhomogeneous steady-state density profiles with an emerging length scale that does not depend on the particle density and is set by $Pe\approx1$. We show that the phenomenological two phase fluid constitutive model is in line with our results.
\end{abstract}

\maketitle

Complex fluids comprise a family of fluids which include elements of intermediate size; elements which are larger compared to the molecules forming the embedding liquid and smaller compared to a typical macroscopic length scale. Among many (e.g. solutions of long polymer chains), suspensions of hard spheres, which we address here, are probably the simplest example. 

Over the years extensive efforts have been invested in exploring the flow behavior of hard sphere suspensions at conditions that are far-from-thermal-equilibrium (i.e non-Brownian suspension) \cite{Denn2014,Guazzelli2018} both experimentally and by modeling. Strong driving significantly modifies the microstructure of the immersed particles \cite{Brady_morris_1997,Cheng_Cohen_Science_2011,Blanc2013}. The coupling between microstructure and hydrodynamic interactions between particles \cite{Brady_morris_1997,Foss2000}, and the frictional contact between the particles \cite{Lin2015,Mari2015}, result in a rich phenomenology which is dramatically different from that of Newtonian fluids. This includes a strong rate dependence of viscosity and the emergence of normal stresses \cite{Brown2014,Guazzelli2018}. The latter, in particular, does not have any equivalence in classical fluids and gives rise to large scale motion (migration) of particles relative to the embedding liquid \cite{Gadala-Maria_Acrivos_1980,Leighton_Acrivos_1987,Morris_Boulay_1999,Boyer_pouliquen_guazzelli_2011,Denn2014,Guazzelli2018}. Resulting large scale variations in the particle density, in turn, affect the overall flow of the composite fluid.

In contrast to the far-from-thermal-equilibrium limit, where the Brownian motion of the particles has negligible contribution, migration at the near-equilibrium regime requires understanding of the interplay between particle flow and their thermal fluctuations. Although the former has been extensively explored in experiments, the study of the later largely relies on numerical simulations. Here we provide measurements of particle migration at both high and low strain rate regimes and directly resolve the spatial crossover between the two. We characterize this crossover across a wide range of driving rates and particle densities, and show that a previously suggested  phenomenological two phase fluid constitutive model \cite{FrankJFM2003} is in line with our observations.

To this end, we adopt a “bottom-up” approach. We probe the response of a colloidal suspension to a local driving on a single particle level; a microscale version of rheometry \cite{Habdas_Weeks_2004,Dullens2011,Pesic2012,Gomez_Solano_Bechinger_2014,Zia_2018}. This is in contrast to the traditional method, focusing on the flow behavior caused by imposing bulk forces on macroscopic samples \cite{Guazzelli2018}.

\begin{figure}
	\includegraphics[width=86mm] {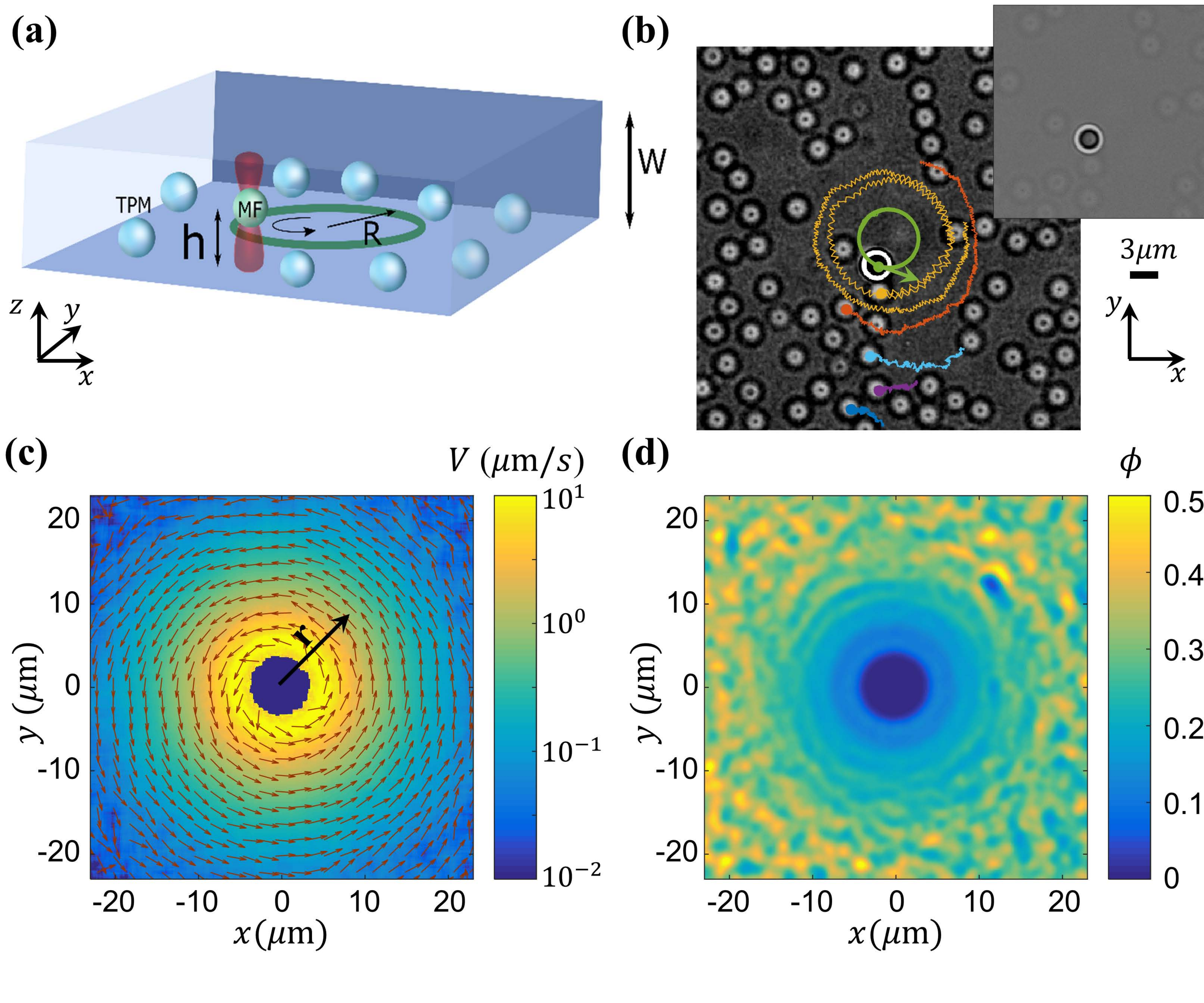}
	\caption{\label{fig1} Steady-state flow and density profiles of a locally driven colloidal suspension. (a) A 2D layer of TPM particles is formed by sedimentation in a sealed cell with thickness w~$\approx170~\mu$m. A MF particle is trapped at $h\approx 2~\mu$m above the coverslip by a focused IR laser beam and prescribed a circular trajectory with a radius $R=3.35~\mu$m (green line). (b) Index of refraction of the solvent and the bath particles (TPM) were chosen to match closely (inset) so that interactions between the laser and TPM particles are negligible. Once the image contrast is enhanced, particle trajectories can be tracked. Five selected trajectories of the bath particles over a period of $20$~s are superimposed. (c) Particle velocity field $V$ reveals azimuthal flow spanning four orders of magnitude within the field of view. Note the logarithmic color axis. (d) Measured local area fraction $\phi$ show a highly non-homogeneous steady-state profile formed by particles migrating from high to low strain rate regions. (b-c) The prescribed driving frequency and equilibrium area fraction are $f=10$Hz and  $\phi_0=0.3$, respectively. Spatial profiles of $V$ and $\phi$ where obtained by temporal averaging of the measurements over a period of $\sim 40 $~min and further smoothed spatially over $0.5 \times 0.5 ~\mu$m$^2$ regions. The trapped particle is excluded in the $V$ and $\phi$ color plots.
	}
\end{figure}

Our experimental system is schematically presented in Fig.~\ref{fig1}a. We sediment two species of spherical colloidal particles to form a quasi-two-dimensional (2D) layer in a range of area fractions $0.1  < \phi_0  < 0.4$. For the majority phase, bath particles, we use 3-trimethoxysilyl propyl methacrylate (TPM) \cite{Liu2016} with a radius of $a=1.35\pm0.05~\mu$m \footnote{Courtesy of Prof. Roel Dullens from the Department of Chemistry, Oxford, GB.}. A Melamine-formaldehyde (MF) particle ($a=1\pm 0.03~\mu$m, Sigma-Aldrich) is trapped above the glass plate and prescribed a circular trajectory by using optical tweezers. The driving frequencies are varied within a range of $0.1  < f <10$ Hz. Self-diffusion of the sedimented bath particles near the bottom solid plane \cite{Sonn-Segev2015,Thorneywork_Dullens_2015} is measured to be $D \approx 0.05~\mu$m$^2/$sec. 

An inverted microscope (Olympus IX71) with a 60$\times$ objective (Olympus, oil immersion, NA=1.42) is used to image the particles and focus the trapping laser beam ($\lambda=1083~$nm). A circular motion of the trapped particle is prescribed by deflecting the laser beam with a piezo tip/tilt platform \cite{Faucheux1995, Gomez_Solano_Bechinger_2014}. Particles are dispersed in an organic solvent mixture of 40/60 (by volume) Tetralin and Decaline that nearly matches the index of refraction, $n$ of the TPM particles  [Fig.~\ref{fig1}(b) - inset]. Since optical trapping depends on the refractive index mismatch between particle and medium, we can manipulate the MF ($n\approx1.68$) particles while barely affecting the TPM particles ($n\approx1.5$) with the laser light. The positions of the TPM  particles can be located and tracked \cite{Crocker1996} by processing the images [Fig.~\ref{fig1}(b)] as a result of a small remaining mismatch in $n$. Commercial dispersant OLOA 1200 is added ($\sim 1\%$ by weight) to tune the electrostatic screening length and prevent particle aggregation and sticking to the coverslip \cite{Liu2016}. Finally, we minimize local heating of the solvent by the focused laser beam and rule out any potential convection of the particles.

A typical example of a steady-state response of the suspension to a local rotational driving is presented in Fig.~\ref{fig1}. Figure~\ref{fig1}(b) demonstrates particle trajectories over a fraction of the experiment duration (see also movie in SM). The prescribed circular trajectory of the driven particle induces an average rotational flow profile of the bath particles with velocity amplitudes that span four orders of magnitude within our field of view [Fig.~\ref{fig1}(c)]. Figure ~\ref{fig1}(d) shows the resulting highly nonuniform spatial profile of the particle density (presented in terms of an area fraction $\phi$) that formed by migration of particles from high to low flow rate regions. This rearrangement of the particles, in the absence of an external potential, is a first confirmation of the strongly out-equilibrium conditions of the suspension. Nevertheless, it is evident that the inhomogeneity in the $\phi(r)$ profile is localized; $\phi \rightarrow \phi_0$ for $r\gg R$. The measured $\phi(r)$ profile, therefore, provides a direct observation of the spatial crossover between the out-of-equilibrium region affected by local driving, and the unaffected equilibrium region. In what follows we characterize particle flow, particle migration and the crossover to equilibrium and compare our measurements with a two-phase constitutive model for colloidal suspensions.

\begin{figure}
	\includegraphics{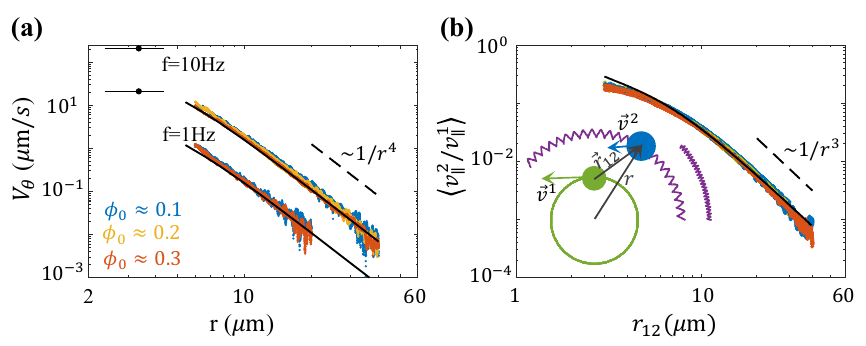}
	\caption{\label{fig2}
		Induced particle flow by a local rotational driving. (a) Average azimuthal particle velocity $V_{\theta}$ at two driving frequencies. Solid lines at the top-left corner denote the velocity and size of the driven particle. (b - illustration) $\vec{v}^{1}$ and $\vec{v}^{2}$ are the driven (green) and bath (blue) particle velocities, respectively. The trajectory of the bath particles (purple) is calculated by integrating the flow field resulting from the prescribed circular motion of the driven particle. Compare the ``sawtooth" trajectories with the measurements presented in Fig.~\ref{fig1}(b). $\vec{r}_{12}$ is the particle separation vector. (b - main figure) Averaged ratio of $v^{1}_{\parallel}$ and $v^{2}_{\parallel}$, parallel velocity components with respect to $\vec{r}_{12}$, provides a measure of the parallel eigenvalue of the mobility tensor $\tilde{B}_{ij}^{12}(\vec{r}_{12})$ (see main text and SM for details). (a,b) Colors denote experiments with different equilibrium area fractions $\phi_0$. Asymptotic solution ($|\vec{r}_{12}| \gg h$ and $r \gg h$) \cite{HappelBrenner1983} and corrected result \cite{DufresnePRL2000} are denoted by dashed and solid lines, respectively.
	}
\end{figure}

We start by quantifying and modeling the induced rotational flow of the particles. Figure~\ref{fig2}(a) demonstrates that the averaged azimuthal particle velocity profiles $V_{\theta}(r)$ are roughly independent of $\phi_0$. Similarly, measurements of the velocity correlation [Fig. \ref{fig2}(b)] and short time self diffusion \cite{Thorneywork_Dullens_2015} show only slight dependence on $\phi_0$ for $\phi_0<0.4$. We neglect, therefore, interactions between the bath particles and model the flow field through hydrodynamic pair interactions between the driven and bath particles that appear in an incompressible viscus liquid at vanishing Reynolds number. The form of the hydrodynamic interactions depends on the spatial confinement of the fluid \cite{Diamant2009}. Here, we consider particle motion at height $h$ above a solid plane. The velocity of particle 2, as a result of a motion of particle 1, is given by $v^2_i=\tilde{B}_{ij}^{12}(\vec{r}_{12};h,a^1)v^1_j$, where $\tilde{B}_{ij}^{12}$ is proportional to the mobility tensor (SM and ~\cite{HappelBrenner1983}). The eigenvalues of $\tilde{B}_{ij}^{12}$, within the leading order in $h/r$, are $\tilde{B}_{\parallel}^{12} \sim 1/r^3$ and $\tilde{B}_{\perp}^{12}\sim 1/r^5$, where $\parallel$ and $\perp$ symbols denote parallel and perpendicular directions relative to $\vec{r}_{12}$, respectively. Once a circular path of particle 1 is prescribed, the trajectory of particle 2 is readily integrated. Such periodic driving results in a ``sawtooth" motion [Fig.~\ref{fig2}(b)] and consistent with our measured trajectories [Fig.~\ref{fig1}(b)]. We obtain an analytical expression for the average particle rotational velocity (SM) which decays as $V_\theta\sim 1/r^4$.

These predictions for $V_\theta$ and $\tilde{B}_{\parallel}^{12}$ describe our measurements asymptotically ($r\gg R$), as shown by the dashed black lines in Fig.~\ref{fig2}(a) and Fig.~\ref{fig2}(b), respectively. A solution for $\tilde{B}_{ij}^{12}$ for all orders in $h/r$ \cite{DufresnePRL2000}, and the corresponding integration of $V_{\theta}$ [SM], provide an excellent description of our measurements at all distances with no adjustable parameters (solid black lines in Fig.~\ref{fig2}(a) and \ref{fig2}(b)). We conclude therefore that the average particle flow can be well approximated by the hydrodynamic pair interactions near a solid plane.

The balance between the shear rate $\dot{\gamma}$ and the Brownian relaxation time $a^2/D$ is characterized by the P\'eclet number $P_e=|\dot{\gamma}|a^2/D$. Measured $V_\theta(r)$ profiles together with $\dot{\gamma}_{\theta r}=r\partial_r(V_\theta/r)$, the only non-vanishing component of the strain-rate tensor in case of a 2D rotational flow \cite{Landau_Lifshitz}, reveal therefore that $P_e (r)\sim 1/r^5$. A unique property of our setup is the ability to spatially resolve a wide range of $P_e$ ($0.1< P_e(r)<100$) and access the near and far-from-equilibrium regimes in a single experiment. 
\begin{figure}
	\includegraphics{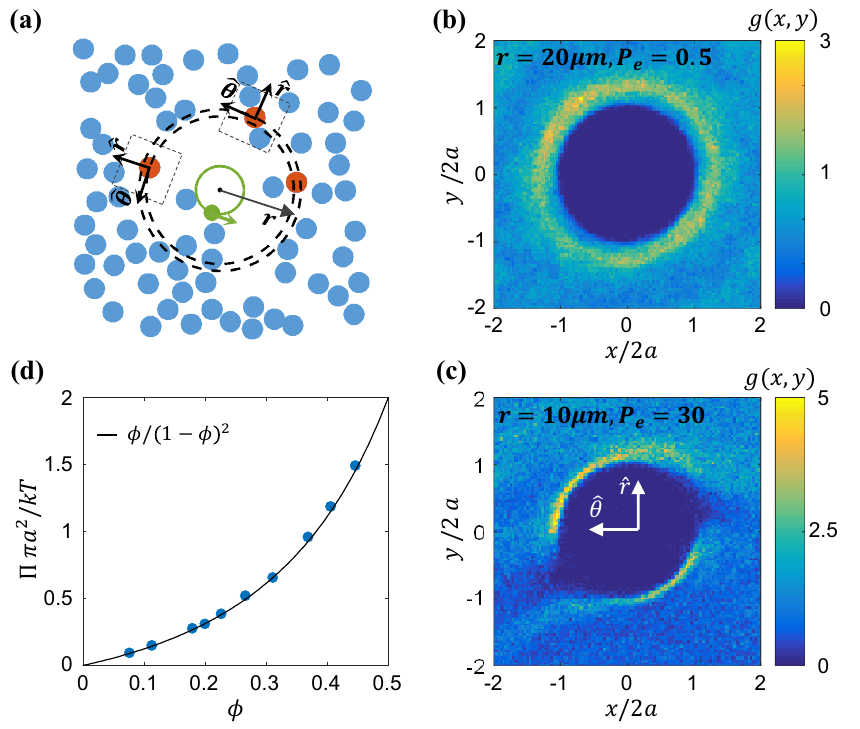}
	\caption{\label{fig3}
		Pair correlation function near and far-from-thermal-equilibrium. (a) For any snapshot in time, as radial gradient in $P_{e}(r)$ are present, $g(x,y)$ is calculated for particles in a narrow ring at distance $r$ from the center (orange particles). Statistics are enhanced by averaging over time and the particles within the ring by using a polar coordinate system $(\hat{r},\hat{\theta})$. Here $x$ and $y$ denote local spatial coordinates in directions of $\hat{\theta}$ and $\hat{r}$, respectively. Measured $g(x,y)$ far from (b) and near (c) the driving region demonstrate the symmetry breaking at far-from equilibrium conditions. $\phi_0=0.3, f=10$Hz, corresponding to the experiment presented in Fig.~\ref{fig1}. (d) At vanishing $P_e$ values osmotic pressure is inferred from measurements of $g$ (blue dots). Theoretical predictions for 2D layer of hard spheres \cite{Helfand1961} is denoted by the black solid line (see main text for more details).
	}
\end{figure}

In Fig.~\ref{fig3} we address the evolved microstructure of our colloidal suspension at both high and low $P_e$ regions by characterizing the 2D pair correlation function $g(\vec{r}_{12})$. In order to account for the spatial variations in $P_e$ we calculate $g(x,y)$ only with respect to the particles that are included in a narrow ring at a distance of interest $r$ from the center of the rotational flow [see Fig.~\ref{fig3}(a) for details]. Here $x$ and $y$ denote the spatial coordinates in the parallel $\hat{\theta}$ and perpendicular $\hat{r}$ directions with respect to the flow. The rotational symmetry and the steady-state conditions of the flow allow us to average $g(x,y)$ over both the particles within the ring and time. The $g(x,y)$ maps obtained far and near the driving, corresponding to low and high $P_e$ regions, are presented in Fig.~\ref{fig3}(b) and Fig.~\ref{fig3}(c), respectively. 

At low $P_e$ regions, where particle diffusion dominates particle advection,  $g(\vec{r}_{12})$ show an isotropic form [Fig.~\ref{fig3}(b)] that indicates an equal probability of finding a particle at any direction with respect to any other particle. At higher $P_e$ values, however, this symmetry is highly distorted [Fig.~\ref{fig3}(c)] \cite{Blanc2013,Cheng_Cohen_Science_2011,Lin_Cohen_PRE_2013,Gao_PRE_2010}. The probability of observing two particles in close vicinity is significantly increased in the second and fourth quadrants. This broken symmetry of $g(\vec{r}_{12})$ reflects the direction of flow. The modified microstructure (local rearrangement) of the immersed particles is a clear signature of nonequlibrium dynamics and a key ingredient in the emergence of non-linear stresses \cite{Brady_morris_1997}. These stresses, in turn, drive the observed spatial variations in $\phi(r)$ on larger length scales [Fig.~\ref{fig1}(d)].

\begin{figure}
	\includegraphics{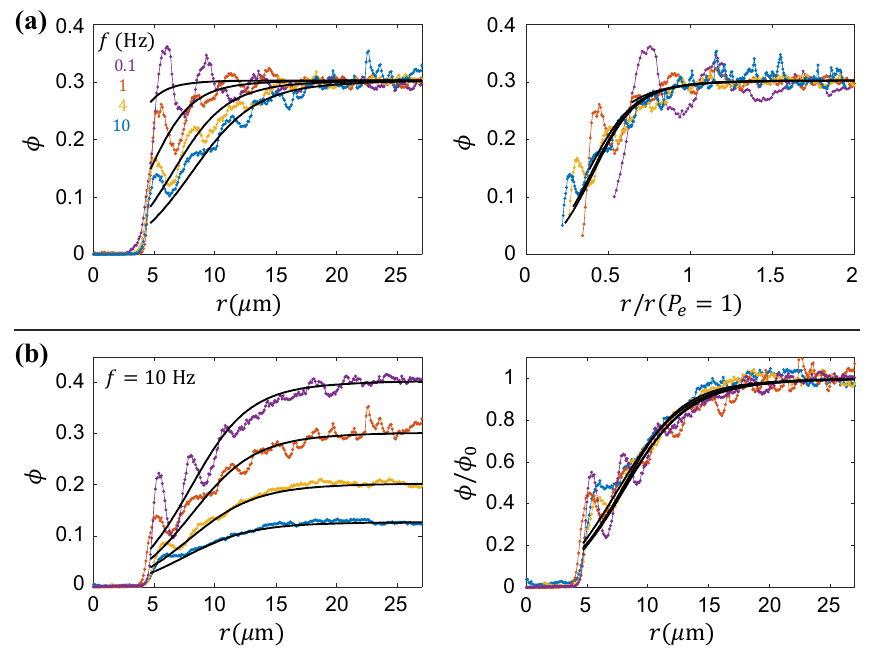}
	\caption{\label{fig4}
		Comparing measured density profiles with rheological constitutive model. (a) Measured radial density profiles, $\phi(r)$ for increasing driving rate. (right) $\phi(r)$ profiles collapse to a single functional form when plotted with respect to $r/r(P_e=1)$, where $r(P_e=1)$ denotes the distance where shearing motion and particle diffusion are comparable. (b) $\phi(r)$ for increasing equilibrium density $\phi_0$. All denisty profiles collapse when scaled by equilibrium values $\phi_0$. (a,b) Solid lines denote predictions of the constitutive model (Eq.~\ref{eq:1}) for $\lambda_{rr}=\lambda_{\theta \theta}=2$.
	}
\end{figure}

We now characterize the the steady state spatial $\phi(r)$ profiles and the crossover to equilibrium. Figure \ref{fig4}(a-left) presents the measured $\phi(r)$ profiles for a wide range of driving rates. Evidently, increasing driving rates amplify particle migration, which in turn results in stronger gradients in $\phi(r)$ and wider particle depletion region. All profiles of $\phi(r)$ collapse to a single functional form [Fig.~\ref{fig4}(a, right)] when $r$ is rescaled by the distance where shear and Brownian motions are comparable $r(P_e=1)\sim (v^1)^{1/5}$. This emerging length scale that characterizes the spatial crossover from far- to near-equilibrium conditions of the suspension is a direct result of the interplay between the flow and thermal fluctuations and is not present in traditional large scale rheometry experiments of non-Brownian suspensions, where particle migration is only limited by the system size \cite{Boyer_pouliquen_guazzelli_2011}. In Fig.~\ref{fig4}(b) we further compare a series of experiments with various equilibrium particle area fractions $\phi_0$. Within the explored range $0.1<\phi_0<0.4$, all profiles of $\phi$ scale with $\phi_0$ [Fig.~\ref{fig4}(b, right)]. This observation demonstrates that the crossover between low and high $P_e$ limits does not depend on the particle density. 

In what follows we compare these observations with a phenomenological two phase fluid model. What are the forces acting on the particles? At vanishing $P_e$ values, Brownian motion of the particles results in an osmotic pressure $\Pi(\phi)$. Any gradients in the particle density will result in a flux of particles $\mathbf{j}_B \sim -\nabla \Pi$. In case of a 2D layer of hard spheres osmotic pressure is related to $g(\vec{r}_{12})$ through $\Pi \pi a^2/kT=\phi[1+2g(2a;\phi)]$, where $g(2a)$ denotes the pair correlation value at particle contact \cite{Brady1993,Lin2016}. We neglect the slight repulsion between the particles, and map our system to the hard sphere case by approximating $g(2a)$ with the maximal $g(\vec{r}_{12})$ value. The inferred values of $\Pi(\phi)$ [Fig.~\ref{fig3}(d)] agree well with the theoretical predictions $\Pi\pi a^2/kT=\phi/(1-\phi)^2$ \cite{Helfand1961} and previous experimental observations \cite{Thorneywork_Dullens_2017}. 

In the presence of strong shearing motion ($P_e\gg1$) the dominant hydrodynamic interactions and asymmetric pair correlation function [Fig.~\ref{fig3}(c)] give rise to normal stresses $\Sigma_{ii}^{\dot{\gamma}}\sim |\dot{\gamma}|$. In the two-phase model, where motion of the particles relative to the embedding liquid is considered, gradients in $\dot{\gamma}$ result in particle flux $\mathbf{j}_{\dot{\gamma}}\sim - \nabla\cdot \mathbf{\Sigma}^{\dot{\gamma}}$, a process often referred to as shear induced particle migration \cite{Morris_Boulay_1999,Guazzelli2018}. Direct measurements of $\Sigma_{ii}^{\dot{\gamma}}$ are challenging \cite{Dbouk2013} and the exact form of the stress components is subject to debate \cite{Boyer_pouliquen_guazzelli_2011,Chu2016,Seto2018,Guazzelli2018}. Furthermore, we are not aware of any existing predictions for the 2D scenario we consider here. Therefore, in order to limit the complexity of the constitutive model, we rely on a simplified stress form predicted for dilute 3D suspensions at $P_e \rightarrow \infty$ and assume $\Sigma_{ii}^{\dot{\gamma}} = a\lambda_{ii} \eta|\dot{\gamma}| \phi^2$ \cite{Brady_morris_1997,Morris_Boulay_1999,FrankJFM2003,Boyer_pouliquen_guazzelli_2011}. Here $\lambda_{rr}$ and $\lambda_{\theta \theta}$ signify the normal stress anisotropy and particle radius $a$ is used to adapt the 3D stress form to the quasi-2D nature of our system.

We model the crossover between the low and the high $P_e$ regimes and the resulting steady-state spatial profiles $\phi(r)$ (Fig.~\ref{fig4}) by balancing the two origins of the radial particle flux  $\mathbf{j}_B+\mathbf{j}_{\dot{\gamma}}=0$ \cite{FrankJFM2003}. In case of a 2D rotational flow, therefore, $\phi(r)$ are obtained by solving
\begin{equation}\label{eq:1}
0=\partial_r\left[\frac{\phi}{(1-\phi)^2} + \frac{1}{6}\lambda_{rr}P_e\phi^2\right] -\frac{1}{6}(\lambda_{\theta \theta}-\lambda_{rr})\frac{P_e\phi^2}{r}
\end{equation}
Here, $P_e(r)$ profiles are inferred from the measured particle velocities [Fig.~\ref{fig2}(a)], whereas $\lambda_{rr}$ and $\lambda_{\theta \theta}$ remain to be determined.

Solutions to Eq.~\ref{eq:1} describe our measurements well in a broad range of driving rates [Fig.~\ref{fig4}(a)] and particle concentrations [Fig.~\ref{fig4}(b)] for $\lambda_{rr}=\lambda_{\theta \theta}=2$. Specifically, the model captures successfully the spatial crossover between the high $P_e$ regime and equilibrium. Any deviations of the model from the scaling in $\phi_0$, as a result of the non-linearity of Eq.~\ref{eq:1}, are beyond the resolution of our measurements.

We note that the particular choice  of $\lambda_{ii}=2$ is motivated by an attempt to maintain simplicity of the model, whereas a strict constraint on $\lambda_{ii}$ proves to be difficult; similar level of agreement between the model and measurements can be obtained in a range  between $\lambda_{rr}\approx3,\lambda_{\theta \theta}\approx1$ and $\lambda_{rr}\approx1,\lambda_{\theta \theta}\approx3$. These values are consistent with other studies in various 3D geometries \cite{Morris_Boulay_1999,FrankJFM2003,Zia_2018}, which show that $\lambda_{rr}$ and $\lambda_{\theta \theta}$ are typically of order unity. 

While the constitutive model captures the average spatial form of the density profiles, it fails to describe their detailed structure. Particle excluded volume gives rise to strong oscillations in $\phi(r)$ (Fig.~\ref{fig4}), which suggests the formation of a layered structure. These particle correlations are increasing with the particle density, but also smeared out by an increasing shear rate. As the model we consider here essentially provides a continuum description of the fluid, density correlations on the particle scale are beyond its' range of applicability. In fact, it's compelling that a continuum model provides such a good description on length scales that are comparable to the particle size. 

One unique feature of our system is that the response of the environment (bath particles) occurs on time scales comparable to the driving rate. This is in striking contrast to traditional colloidal model systems studied in the context of stochastic thermodynamics out of equilibrium \cite{Seifert2012,Blickle2006,Berut2014}. We suggest, therefore, that our mean-field continuum analysis provides a starting point to explore various out-of-equilibrium statistical mechanics observables such as self- and pair-diffusion, entropy production and fluctuations in driven systems where bath dynamics can not be neglected. 

\begin{acknowledgments}
We thank Haim Diamant for insightful discussions and for deriving the analytical expression for the average rotational flow. We thank Yanyan Liu and Roel P. A. Dullens for synthesizing the TPM particles. We acknowledge funding from the Israeli Science Foundation grant no: 988/17
\end{acknowledgments}

%\bibliography{ref,ISF15}

\end{document}